\documentclass[
twocolumn,
]{revtex4-2}
\usepackage[T1]{fontenc}

\usepackage[
    a4paper, 
    mag=1000,
    left=1.5cm, 
    right=1.5cm,
    top=2cm,
    bottom=2cm,
    headsep=0.7cm,
    footskip=1cm
    ]{geometry}
    
\usepackage[utf8]{inputenc}
\usepackage{amssymb}
\usepackage{mathtools}
\usepackage[T2A]{fontenc}
\usepackage{amsmath}
\usepackage[russian,english]{babel}
\usepackage{graphicx}
\usepackage[table]{xcolor}

\usepackage{amsfonts} 

\usepackage{enumerate}

\usepackage{graphicx}
\usepackage{dcolumn}
\usepackage{bm}
\usepackage{hyperref}
\usepackage{array}
\usepackage{comment}
\usepackage{float}
\usepackage{subcaption}
\usepackage{xcolor}

\begin{document}

\title{Trainable dynamical masking for readout-free optical computing}

\author{S. Bogdanov*, E. Manuylovich and S. K. Turitsyn} 

\affiliation{Aston Institute of Photonic Technologies, Aston University, Birmingham B4 7ET, UK}
\email{s.bogdanov@aston.ac.uk}

\begin{abstract} 
Nonlinear systems, transforming an input signal into a high-dimensional output feature space, can be used for non-conventional computing. This approach, however, requires a change of system parameters during training rather than coefficients in a software program. We propose here to use available off-the-shelf high-speed optical communication devices and technologies to implement a trainable dynamical mask in addition to or even instead of the traditional readout layer for extreme learning machine-based computing. The computational potential of the proposed approach is demonstrated with both regression and time series prediction tasks. 
\end{abstract}

\maketitle

\section{Introduction}

Reservoir computing (RC)~\cite{RC01, RC02} and extreme learning machine (ELM)~\cite{ELM01} are two widely used non-digital computational frameworks suited for sequential (e.g., temporal) data processing. The  RC architectures exploit recurrent connections, creating memory in the system, while ELM is a feed-forward approach. In both scenarios, an input signal is nonlinearly transformed into a high-dimensional output that can be processed using simple and efficient linear algorithms. 
The projection onto a higher-dimensional output space allows for the separability of features that are indivisible in the original input signal. 
Nonlinear transformation can be implemented using various physical systems, substrates, and devices \cite{Jaeger2023,Tanaka_2019,Wright_2022,Logan02}. 
This enables the development of a new platform for analog machine learning,  artificial neural networks, and non-conventional computing based on the existing physical effects and systems rather than on specially designed computing devices. Note that the idea that nonlinear physical systems can be harnessed for computation is conceptually close to the seminal Feynman's paper "Simulating physics with computers" \cite{feynman1982simulating} that established a foundation for quantum computing. 

In particular, wave physics can be used for analog processing of time-varying signals \cite{WAves01,marcucci2020theory,Jalali01,SC01}. 
Evolving time dynamics of the wave equation resembles the operation of recurrent neural networks \cite{WAves01}. Physical wave systems in different fields, ranging from nonlinear optics and hydrodynamics to polaritonics and Bose-Einstein condensates, can be trained to learn complex features in temporal data. Coherent structures and highly nonlinear physical effects, including supercontinuum generation, solitons, breathers, rogue, and shock waves, can be employed to perform computation \cite{marcucci2020theory,SC01}. In this framework, the recurrence relationships occur naturally in the time dynamics of the physical systems, and the propagating waves provide the memory and capacity for information processing \cite{WAves01}.
Effectively, this approach offers analog computing platforms based on the natural nonlinear evolution of a continuous physical system and can be treated as a variant of the ELM scheme that both takes advantage of a feed-forward approach and memory created by wave dynamics. One of the important challenges in implementing physics-based neural networks is the availability of an efficient means of changing a number of physical parameters of the system during training. 
Therefore, operating only with a linear readout layer is beneficial for practical application compared to manipulating distributed system parameters during propagation. 

Implementing physics-based RC and ELM in the optical domain offers advantages such as a high speed of signal processing and a wide variety of physical phenomena to create a reservoir for computing \cite{solli2015analog,mcmahon2023physics,SC01}. In general, nonlinear and multimode fibers, microresonators, semiconductor devices, photonic integrated circuits, random media, topological devices, and metasurfaces can all be used to develop optical accelerators \cite{marcucci2020theory,dynamics,oguz2024programming,disorder,Jalali01,SC01,xu202111,meta} and new computational paradigms. Moreover, a well-developed fiber-optic communications environment can provide practical solutions integrated into compact, commercially available high-speed, high-bandwidth devices \cite{manuylovich2025optical}. Optical and optoelectronic telecom-grade devices are suitable for performing computing tasks. For example, the nonlinear properties of a semiconductor optical amplifier and nonlinear optical loop mirror have been used to demonstrate reservoir computing and extreme learning machine paradigms \cite{manuylovich2024soa,manuylovich2025optical}.

Feedforward neural networks can approximate complex nonlinear mappings directly from the input signal. 
However, the learning speed of such networks is generally slow due to the use of gradient-based learning algorithms to train neural networks and the need to tune all the parameters iteratively.
Physics-based ELM allows us to take advantage of feedforward approaches without the need for backpropagation because such systems usually have a small number of parameters and can be trained with a wide range of optimization techniques. However, the main technical challenge is the practical implementation of the training process that requires changes in the real system's physical parameters. Thus, the speed of training is determined by the speed of the efficient changes of these amenable system parameters. The computational ability of a particular physical system can be gained by using dynamical trainable masking, an additional modulation of the signals carrying information aimed to exploit a nonlinear transformation of an input signal to the high-dimensional output.
This approach has been demonstrated with delay-based RC systems \cite{appeltant2014constructing, kuriki2018impact}.
Dynamical masking does not require cumbersome manipulation of the system's inner physical parameters, relying on feedback to adjust trainable masks.
A similar technique has been used for spatial masking, where a spatial light modulator is applied to the input signal \cite{oguz2024programming}, demonstrating the applicability of the method beyond temporal modulation.
In this Letter, we propose the scheme of extreme learning machine-based computing with a high-speed dynamical trainable mask based on a standard telecom Mach-Zehnder modulator (MZM) that can enhance or even replace the
traditional readout layer. This can dramatically simplify the hardware implementation of the scheme in comparison to the conventional high-dimensional readout layer.

\section{Basic concept and mathematical model}

\begin{figure}[H]
    \centering
    \includegraphics[width=0.99\linewidth]{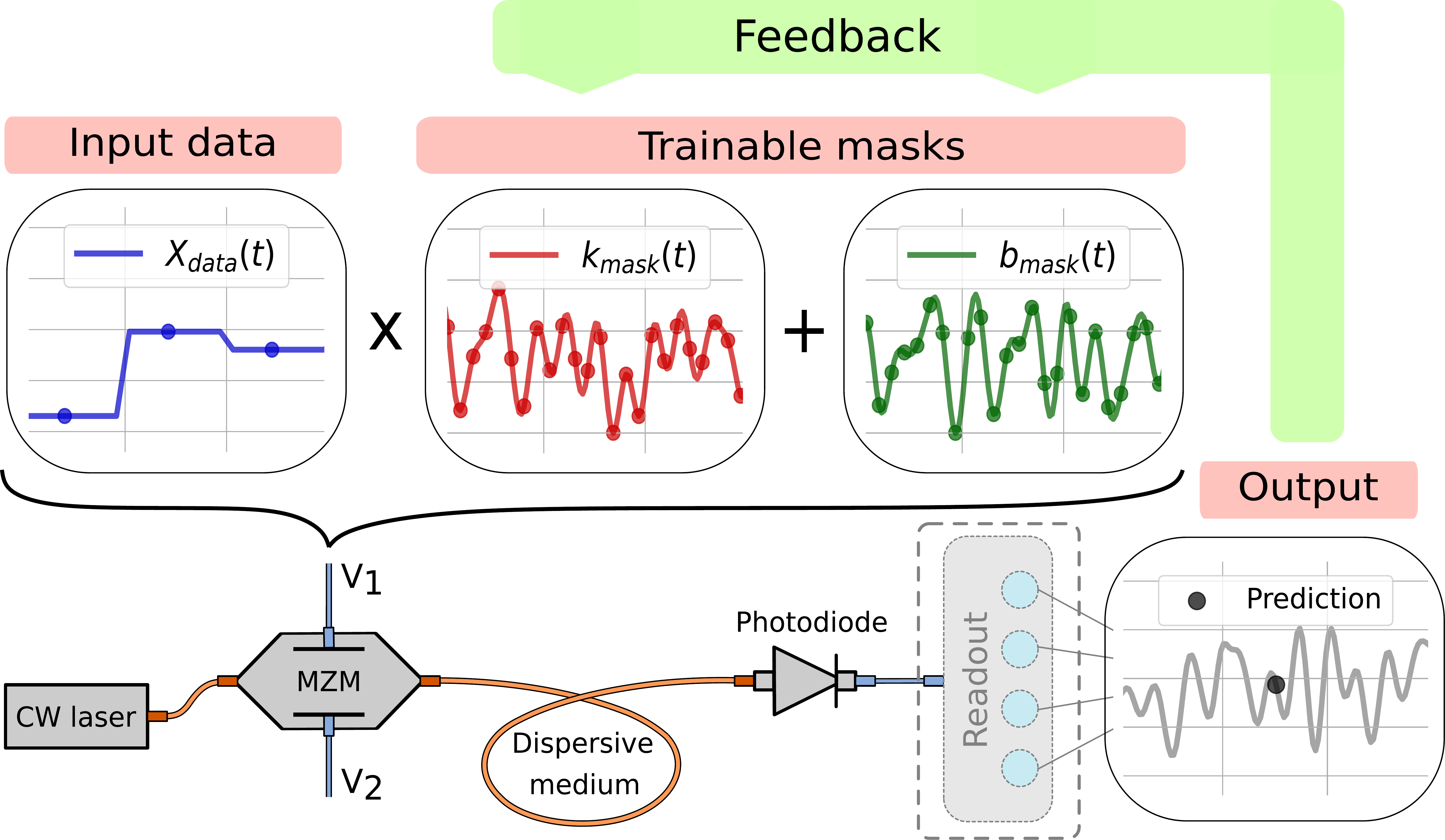}
    \caption{The principal scheme of the computing system with MZM, dispersion medium, photodetector, and optional readout layer. The upper frames depict input data and both trainable masks: multiplicative and biased. The output frame shows the prediction based on a single photocurrent sample.}
    \label{fig:system_proc_z}
\end{figure}

We introduce a simple computation system based on (i) a Mach-Zehnder modulator, (ii) a linear dispersive medium (DM) (e.g., a piece of optical fiber or a chirped fiber Bragg grating), (iii) a photodiode, and (iv) a readout layer that can be removed, see Fig.~\ref{fig:system_proc_z}. The input information stream and trainable dynamical masks modulate the governing voltage on the MZM's arms. The masks being learned from data become shaped to exploit the nonlinear transfer properties of MZM in the most efficient way for the problem under consideration. The optical input of MZM is a continuous wave (CW) from a laser. The dispersive medium introduces additional data mixing (recurrence). The photodiode converts the signal to the electrical domain, where it is sampled and collected into the input vector of the readout layer that can be shrunk to the size of one (readout-free model). We consistently train the coefficient of the mask and weights of the regression layer through a unified process. The evolution of the signal propagated through the elements of the proposed system can be described as a chain of the following transformations.
The Mach-Zehnder modulator transforms the input optical signal according to the applied governing voltage that carries the data and the trainable modulation:
    \begin{equation}
        E_{MZM}^{out}(t) = T_{MZM}(V_1,V_2) \times E_{MZM}^{in}(t),
    \end{equation}
    with the standard transfer function:
    \begin{equation} \label{eq:mzm-res}
        T_{MZM}(V_1,V_2) = cos\left(\frac{\pi (V_1 - V_2)}{2V_{\pi}} \right) e^{\frac{i \pi (V_1 + V_2)}{2V_{\pi}}},
    \end{equation}
 where $V_1$ and $V_2$ governing voltages on two MZM's arms. The transfer function is factorized, enabling modulation of amplitude and phase separately. When $V_2=-V_1$, the MZM operates in so-called "push-pull" mode and modulates the amplitude of the input signal only. $V_{\pi}$ is the switching value such that at the voltage difference $V_1 - V_2 = V_{\pi}$ the modulator is fully closed (see details e.g. in \cite{winzer2008advanced, souza2017analytical}). In simulations, we restrict the bandwidth of the governing voltage signal with a $45\, \mathrm{GHz}$ cut-off frequency filter, mimicking the realistic behaviour of devices.
            
The MZM's output signal $E_{MZM}^{out}(t)$ then propagates through the dispersion medium (e.g., a piece of optical fiber), with the transfer function that has a well-known form in the frequency domain:
\begin{equation}
    E_{DM}^{out}(\omega) = E_{DM}^{in}(\omega) e^{-\frac{i}{2} \omega^2 \, A },
\end{equation}
where $A = D\,\lambda^2/(2\pi c)$, $D$ is the total accumulated dispersion in $\mathrm{ns/nm}$, $\omega$ is circular frequency, $\lambda$ is a central wavelength of the signal, and $c$ is speed of light \cite{agrawal2000nonlinear}.

The signal after the dispersive medium is detected with the photodiode:
\begin{equation}
    I_{PD}^{out}(t) = \kappa \; \int_{-\infty}^{+\infty} H(\tau) |E_{PD}^{in}(t-\tau)|^2 d\tau.
\end{equation} Here $I_{PD}^{out}(t)$ is the photodiode output current, $\kappa$ is photodiode responsivity, $E_{PD}^{in}(t)= E_{DM}^{out}(t)$ is the input optical field, and $H(\tau)$ is photodiode's impulse response function \cite{bowers2003ultrawide}.

Finally, at the readout, $N_{readout}$ subsequent photocurrent samples are multiplied element-wise by a trainable vector of weights of the same size, summarized, and biased with DC forming the predicted value. However, our system can operate without readout having only a single element in the vector of weights.

\section{Dynamical masking}

The voltage $V_1$ is used to seed the data into the model and to incorporate a mask providing optimal coding. The MZM operates in the balanced driving mode when $V_2=-V_1$, and according to Eq.~(\ref{eq:mzm-res}), an input optical signal experiences amplitude modulation only. We map the data stream and mask onto the governing voltage by the following relation:
\begin{equation} \label{eq:voltage}
    V_1(t) = V_{\pi} \Big( k_{mask}(t) \times X_{data}(t) + b_{mask}(t) \Big).
\end{equation}
$X_{data}(t)$ is an input data sequence consisting of $M$ symbols. $k_{mask}(t)$ and $b_{mask}(t)$ are multiplicative and additive trainable masks, each having the size of $N_{mask}$ samples per symbol. Therefore, for given $M$ input symbols, there are $M\times N_{mask}=N_{out}$ independent values for $k_{mask}(t)$ and $b_{mask}(t)$. $V_{\pi}$ is the switching voltage from Eq.~(\ref{eq:mzm-res}). Thus, the proposed system constructed from the available telecom components is capable to nonlinearly transform input $M$-dimensional vector into $M\times N_{mask}$-dimensional output.

The key property of the proposed scheme is high-speed dynamical trainable masking. ``Dynamical'' here means that the mask is a function of time that modulates the input data sequence at the MZM speed. We use the term ``trainable'' to emphasize that the values of the mask are adjusted through the supervised learning process. From this point of view, the system can be considered as a simple neural network consisting of MZM-layer with trainable weights ($k_{mask}(t)$ and $b_{mask}(t)$ in governing voltage, see Eq.~(\ref{eq:voltage})), DM-layer, photodiode-layer both transferring the signal and trainable $N_{readout}$-dimensional readout layer. All trainable parameters are adjusted consistently using a gradient descent algorithm implemented with the TensorFlow 2.0 framework. We observed that the initial values of $k_{mask}(t)$ and $b_{mask}(t)$ that are set to be random need to be scaled by a factor of 5 each. This helps the algorithm to converge to a local minimum in the area where MZM demonstrates a moderate degree of nonlinearity. In other words, we move the initial values of the mask before training to the desirable area. We analyze the performance of the proposed scheme on two basic tests: regression and time series prediction.

\section{Regression task}

\begin{figure}[t]
    \centering
    \begin{subfigure}[b]{0.51\linewidth}
        \includegraphics[width=\linewidth]{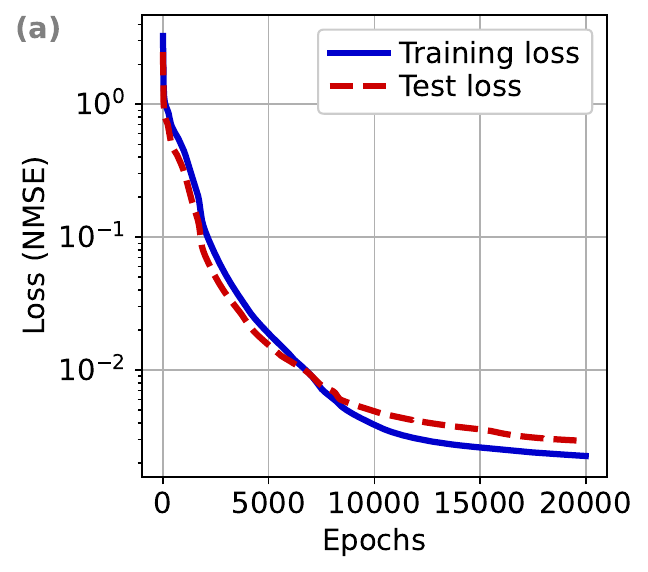}
    \end{subfigure}
    \begin{subfigure}[t]{0.47\linewidth}
        \includegraphics[width=\linewidth]{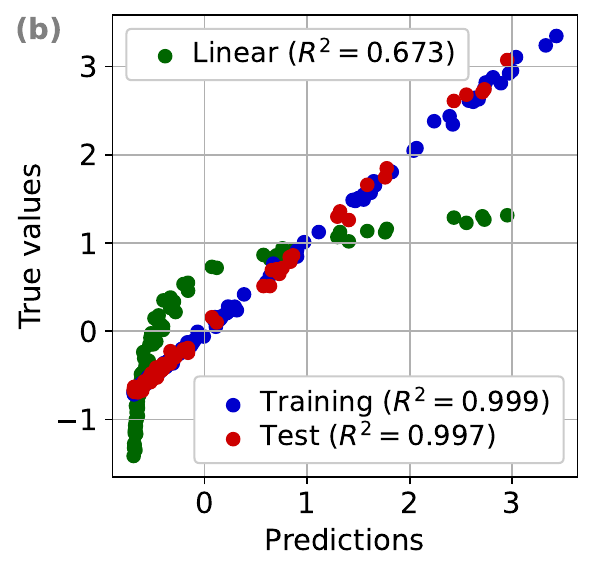}
    \end{subfigure}
    \caption{(a) Training and test loss and (b) prediction results for the readout-free and linear regression model.}
    \label{fig:yacht_data_example}
\end{figure}

We use the yacht hydrodynamics prediction test, which is a simple regression problem to estimate the residuary resistance of a yacht for its given geometric parameters and Froude number. The dataset consists of 308 measures, each of 6 features and one target variable \cite{yacht_hydrodynamics_243}. We use the benchmarking results of \cite{baressi2019artificial}, where the residuary resistance predictions for the same data were performed with a multilayer perception regressor. The authors reported that only 116 models demonstrated performance with a coefficient of determination satisfying $\mathrm{R^2}>0.97$ among 2304 different architectures.
The best model had four hidden layers with the number of neurons 10, 20, 20, 10 correspondingly and $\mathrm{R^2}=0.99329$. It is worth noting for further reference that such a model has 870 trainable parameters (where the connections of the input and output layers are countered).
Also, for comparison, we apply a linear regression model (LRM) for these data to estimate the performance of linear techniques. The data were normalized to zero mean value and variance of one and separated into training and testing parts in proportion $70\%$ and $30\%$, respectively. The ridge regression was trained to predict the target variable with the regularization parameter optimized with the 5-fold cross-validations method. The performance of the LRM is $\mathrm{NMSE}=0.2738$ and $\mathrm{R^2} = 0.6727$. The peculiarity of the mapping from input variables to target value is its nonlinearity. The linear regression model fails to approximate it; see Fig.~\ref{fig:yacht_data_example}.

We implement the predictions of the residuary resistance with our readout-free model based on MZM and demonstrate that our scheme outperforms linear techniques. We train the masks and readout layer using $70\%$ of data and keeping other $30\%$ for the test. Data were normalized to have zero mean and maximum absolute value equal to one to control the maximum voltage at the MZM arms. 

\begin{figure}[b]
    \centering
    \begin{subfigure}[b]{0.48\linewidth}
        \includegraphics[width=\linewidth]{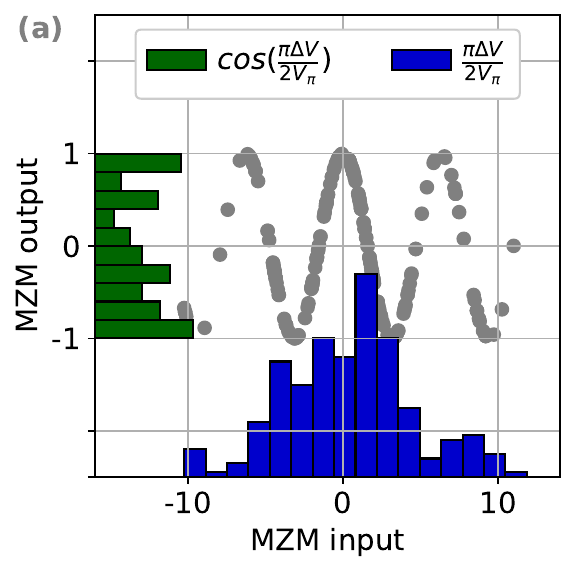}
    \end{subfigure}
    \begin{subfigure}[b]{0.5\linewidth}
        \includegraphics[width=\linewidth]{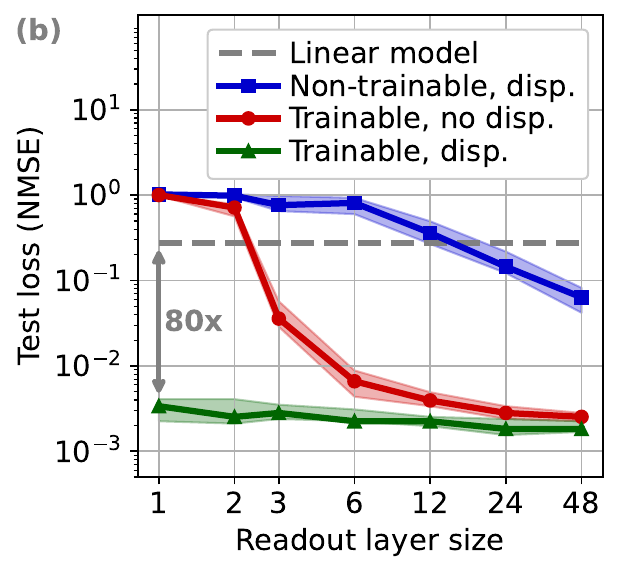}
    \end{subfigure}  
    \caption{(a) The distribution of the argument $\frac{\pi \Delta V}{2 V_{\pi}}$ of the trained model and corresponding to it the output of MZM transfer function, Eq.~(\ref{eq:mzm-res}). (b) The dependence of the model's effectiveness on the number of elements for varying readout sizes 8.}
    \label{fig:mask_and_readout}
\end{figure}

The result of the particular implementation is provided in Fig.~\ref{fig:yacht_data_example}, where we depict training and test loss and the final distribution of the predicted values of the normalized residuary resistance. In this example, the following parameters are chosen: the mask's size is eight elements per symbol, providing a total length (for six symbols) of 48 independent trainable parameters in each multiplicative and additive mask. The dispersion parameter is $D=0.4 \, \mathrm{ns/nm}$. We compare the received predictions with the linear regression result. Our model can effectively approximate the nonlinear mapping between input features and the target variable while the linear technique fails.
Fig.~\ref{fig:mask_and_readout}a depicts the distributions of the MZM input that is $\frac{\pi \Delta V}{2V_{\pi}}$, where $\Delta V = V_1 - V_2$ and its output $\cos\left(\frac{\pi \Delta V}{2V_{\pi}} \right)$ according to the transfer function, Eq.~(\ref{eq:mzm-res}). The distributions are shown for the same model implementation with results depicted in Fig.~\ref{fig:yacht_data_example}.

The main result of this paper is presented in Fig.~\ref{fig:mask_and_readout}b, where we show the model's performance in dependence on the readout layer size (here we deviate from the readout-free baseline model). Three different configurations are considered: (i) trainable masks with dispersion, (ii) trainable masks without dispersion (the dispersion element is removed, see Fig.~\ref{fig:system_proc_z}), and (iii) non-trainable (random and fixed) masks with dispersion, see Fig.~\ref{fig:mask_and_readout}b. This result reveals the critical role of mask training and the presence of dispersion.
For the model with adjustable masks and dispersion, we demonstrate that reducing the number of weights on the readout up to one (readout-free configuration) only minimally degrades performance. In this scenario, the function of trainable readout is effectively transferred to the masks.
Moreover, dispersion becomes a key factor when the readout size is smaller than the number of input features. For the considered task with six input features, reducing the readout size to 3, 2, or even 1 provides a loss of information. The role of dispersion here is to bring this information from other parts of the signal. To plot Fig.~\ref{fig:mask_and_readout}b, we run the model training 25 times for each configuration and depicted the median values and range from $25\%$ to $75\%$ percentiles.

This principally nonlinear problem has been chosen to demonstrate the property of the MZM-based computing system. Even though the proposed model contains a maximum of 96 trainable parameters for two masks of 48 elements each (8 elements per symbol) and two elements at the readout, it provides better performance compared to the multilayer perception regressor model proposed before while having an order of magnitude smaller the number of trainable parameters.

\section{Mackey-Glass time series prediction}

The following differential equation describes the Mackey-Glass (MG) time series \cite{mackey1977oscillation}:
\begin{equation}
    \frac{dx(t)}{dt} = \frac{\beta x(t-\tau)}{1+x^n(t-\tau)} - \gamma x(t),
\end{equation}
where $\beta$, $\gamma$, $n$, and $\tau$ are real-valued parameters.
The $x(t)$ function is known for every $t$ if MG parameters and the initial conditions $\{x(t), t \in (-\tau, 0]\}$ are specified. 
In our research we set $\beta=0.2$, $\gamma=0.1$, $n=10$, $\tau=17$ that are popular values in literature. 
The Mackey-Glass time series is an example of a chaotic system, and its prediction is a classical benchmark problem. Previously, the ability of a reservoir computing system to make such predictions was considered \cite{RC01}. For the same parameters, it was reported $\mathrm{NRMSE_{84}}=0.00028$ and $\mathrm{NRMSE_{84}}=0.00012$ for 300 and 21000 steps of training sequence correspondingly (the index indicates the error of the $84th$ step prediction). Later, the same prediction problem was solved with a reservoir computing implemented based on a semiconductor optical amplifier \cite{manuylovich2024soa}. The prediction accuracy achieved in this work was $\mathrm{NMSE_{300}}=0.01$.

\begin{figure}[H]
    \centering
    \begin{subfigure}[b]{0.475\linewidth}
        \includegraphics[width=\linewidth]{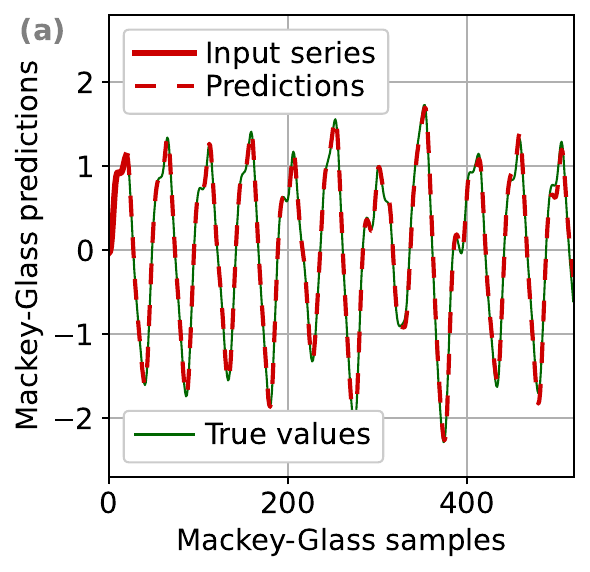}
    \end{subfigure}
    \begin{subfigure}[b]{0.51\linewidth}
        \includegraphics[width=\linewidth]{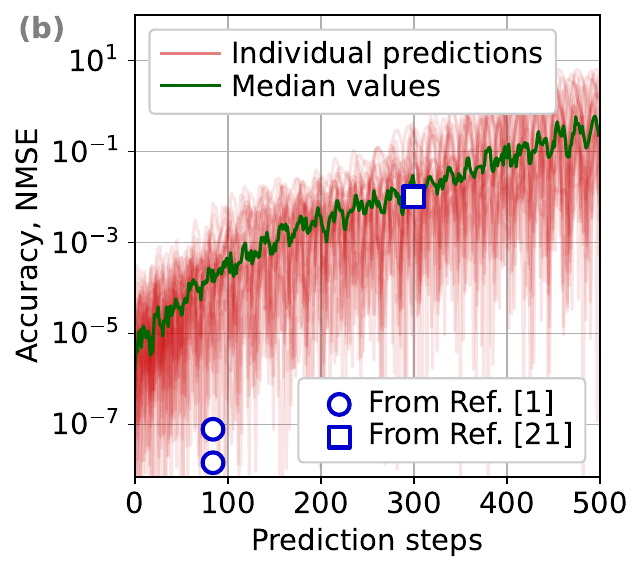}
    \end{subfigure}  
    \caption{(a) The Mackey-Glass time series prediction with the readout-free model. (b) NMSE for different prediction steps. 
    For comparison, the results from \cite{RC01} and \cite{manuylovich2024soa} are depicted.}
    \label{fig:MG_8}
\end{figure}

We performed the MG series prediction with our readout-free model with the same mask size (8 elements per symbol) and dispersion as in the previous example. However, in this task, the number of input symbols is 20, providing 
160 trainable parameters in each mask. We report Mackey-Glass time series prediction up to 500 steps; see Fig.~\ref{fig:MG_8}a. We also plot $\mathrm{NMSE}$ as a function of the prediction step and compare it with the result from \cite{RC01, manuylovich2024soa}. In Fig.~\ref{fig:MG_8}b, there are the results of 25 independent predictions and corresponding median values calculated for each prediction horizon independently.

\section{Discussion and conclusions}

We designed and demonstrated via numerical modeling a new physics-based ELM computational scheme using commercially available high-speed optical communication components and technologies. The main novel elements of the proposed approach are: (i) applying a well-developed optical communication technique of temporal up-sampling of the input signal to enable mapping of data onto high-dimentional feature space, (ii) using a standard high-speed Mach-Zehnder modulator as a core element of computing system for trainable dynamical masking of the input signal, (iii) demonstrating that the model with the dynamic masks could operate without a traditional readout layer. We believe this work might pave the way for new unconventional computing schemes using various high-speed telecom-grade devices and technologies. 

\section{Acknowledgment} We acknowledge support by the Engineering and Physical Sciences Research Council (project EP/W002868/1). 

\bibliography{sample}

\end{document}